\documentclass[epsfig,psfig]{aa}

\usepackage{graphicx}
\usepackage{txfonts}
\begin{document}

\title{Are dry mergers of Ellipticals the way to reconcile model predictions with the downsizing?}
\titlerunning{Dry mergers in ellipticals?}

\author{Antonio Pipino$^{1}$ \& Francesca Matteucci$^{2,3}$}
\authorrunning{A. Pipino \& F. Matteucci}

\institute{$^1$ Astrophysics, University of Oxford, Denys Wilkinson Building, Keble Road, Oxford OX1 3RH, U.K\\
$^2$ Dipartimento di Astronomia, Universita di Trieste, Via G.B. Tiepolo, 11, 34100 Trieste, Italy\\
$^3$ INAF- Trieste, Via G.B. Tiepolo 11, 34100 Trieste, Italy}
\date{Accepted,
      Received }

%\maketitle

\abstract{}
{To show that the bulk of the star formation and the galaxy assembly should occur simultaneously
in order to reproduce at the same time the \emph{downsizing} and the chemical properties of present-day
massive spheroids within one effective radius.}
{By means of chemical evolution models we create galactic building blocks of several masses
and different chemical properties. We then construct a sample of possible merger histories
going from a multiple minor merger scenario to a single major merger event aimed at reproducing
a single massive elliptical galaxy.
We compare our results against the mass-[Mg/Fe] and the mass-metallicity relations.}
{We found that a series of multiple dry-mergers (no star formation in connection with the merger) involving building-blocks which have been created ad hoc in order to satisfy
the [Mg/Fe]-mass relation cannot fit the mass-metallicity relation and viceversa.
A major dry merger, instead, does not worsen the agreement with observation if it happens
between galaxies which already obey to both the mass($\sigma$)-[Mg/Fe] and the mass($\sigma$)-metallicity
relations. However,
this process alone cannot explain the physical reasons for these trends.} 
{Dry mergers alone cannot be the way to reconcile the need of a more efficient
star formation in the most massive galaxies with the late time assembly 
suggested in the hierarchical paradigm
in order to recover the galaxy downsizing.}

\maketitle

\keywords{galaxies: elliptical and lenticular, cD - galaxies: abundances - galaxies: formation - galaxies: evolution}

\section{Introduction}

{ The picture for the formation of elliptical galaxies
in the framework of the hierarchical clustering scenario, namely of a major merger involving
two spirals at late time (e.g. Kauffmann \& White 1993) has been questioned
several times (e.g. Ostriker 1980) since its original formulation.
Recent studies (e.g Thomas \& Kauffmann 1999; Pipino \& Matteucci 2006, Naab \& Ostriker 2007),
emphasised a tension between the observed photo-chemical properties and the predicted ones.
From the dynamical view-point, however, the situation is different. In fact,
at least the medium-sized fast-rotating ellipticals (Emsellem et al. 2007) have 
global morphological and kinematical properties which resemble
those of a spiral-spiral merger remnant (e.g. Naab \& Burkert 2003, Cox et al. 2006, Robertson et al. 2006).
On the other hand, the most massive ojects are better represented by the outcome
of a dissipationless merger (Naab et al., 2006).

In a more general perspective}, in order to reconcile at the same time the anti-hierarchical behaviour of the AGNs (e.g. Hasinger et al. 2005), 
the evolution of luminosity function with redshift (e.g. Bundy et al. 2006) as well as the
evidences coming from the analysis of the stellar populations inhabiting ellipticals (e.g.
Thomas et al. 2002), a substantial modification of the baryons behaviour,
whith respect to the Dark Matter one, seems to be required.
In particular, more massive ellipticals are older and formed faster with
respect to smaller objects (Thomas et al. 2005).
This is the so-called \emph{downsizing} (Cowie 1996). The well known downsizing in the \emph{chemical}  
properties of ellipticals, namely the increase of mean stellar [Mg/Fe] with galaxy mass (see Matteucci 1994), has
received further independent evidences supporting it. 
For instance, the very recent observations of the evolution of the mass-gas metallicity relation with redshift (Maiolino et al. 2008)
and the study of the present-day ratio between stars and gas 
(Calura et al. 2007), both
favour the earlier and faster completion of the SF process for the most 
massive spheroids, with respect to the low-mass ones.

Hierarchical modelling, in its latest versions, partly accounts for
the downsizing. In pratice, the mass assembly still occurs at late times,
but most of the stars have been formed at high redshift in small subunits.
The preferred mechanism for the assembly of massive spheroids
is a sequence of dry-mergers\footnote{In this paper \emph{dry-merger} means a pure dissipationless
merger of stellar systems, i.e. without any gas and star formation} (e.g. De Lucia et al. 2006).
Dry mergers have been observed (Tran et al. 2005, Bell et al. 2006, Rines et al. 2007; and in more extended samples by van Dokkum 2005),
although the criteria used to observationally define a dry-merger have
been questioned by, e.g. Donovan et al. (2007).
%The inference of dry-merger rate at several epochs is even a more
%difficult task.
%Direct observation of mergers pairs seems to suggest..
Dry merger between spheroidal systems are also invoked to explain
the so-called boxy ellipticals (e.g. Naab et al. 2006).
However, if we restrict ourselves to the most massive elliptical galaxies ($ L > L_*$),
they seem to be in place and do not show any signs of significant evolution
in mass since $z \sim 1$ (Scarlata et al. 2006, Brown et al. 2007).
A great deal of work has been done in the field of dynamical simulations,
but so far the consequences on the chemical properties
of the final stellar populations have not been tested.

On the other hand, in the \emph{revised monolithic} scenario (Larson, 1974, Matteucci, 1994; Chiosi \& Carraro
2002; Merlin \& Chiosi 2006) both the mass(or $\sigma$)-[Mg/Fe] (MFMR, hereafter) and the mass(or $\sigma$)-metallicity (MMR)
relations are naturally explained,  as shown by
Pipino \& Matteucci (2004, PM04 hereafter).
In particular, in PM04 for the first time the inverse
wind scenario (Matteucci 1994) plus an initial infall episode are adopted whithin a multi-zone
formulation. 
Under reasonable assumptions on the behaviour of the infall timescale
and the star formation (SF) efficiency with galactic mass, PM04
showed how this kind of model can reproduce
the whole set of chemical and photometric observables simultaneously.

\emph{In particular, it is necessary that both the bulk of the star formation
and the galactic assembly proceed in lockstep}.
The same conclusion was reached by Cimatti, Daddi \& Renzini (2006), who showed that the downsizing trend should be extended also to the mass assembly, in the sense that the most massive ellipticals should have assembled before the less massive ones). This conclusion was based on a re-analysis of the rest frame B-band COMBO-17 and DEEP2 luminosity functions.
The aim of this paper is to show that this has not been taken
yet into account in models based on the hierarchical clustering paradigm.
We create galactic building blocks of different mass
and chemical properties. We then construct a sample of possible merger histories
running from a multimple minor merger scenario to a single major merger event aimed at reproducing
a single massive elliptical galaxy.
We compare the results againts the MFMR and the MMR.
In the absence of full a dynamical treatment, we cannot undertake a deeper 
analysis of the
mass- and [$<Mg/Fe>$]-$\sigma$ relations. In the following we will refer
to the stellar velocity dispersion as a mass tracer, unless otherwise stated.

\section{The model}

The chemical code adopted here is described 
in PM04, where we address the reader for more details. In particular,
this model is characterized by:
Salpeter (1955) IMF, Thielemann et al. (1996) yields for massive stars,
Nomoto et al. (1997) yields for type Ia SNe and 
van den Hoek \& Groenewegen (1997) yields for low-
and intermediate-mass stars (the case with $\eta_{AGB}$ varying with metallicity). 

We will use this model for producing ad hoc progenitors of present-day galaxies
in order to investigate whether the final composite stellar population (CSP) in the merger remnant has
properties which match those of observed ellipticals.

In particular, for each of them, we list star formation efficiency, infall timescale,
and average stellar properties, such as the mass-weigthed abundances and abundance ratios
(see PM04 for their definition). 
%We deal with the baryonic content of the galaxies,
%therefore the masses always refer to the amount of \emph{luminous} matter.

We define, according to Pagel \& Patchett (1975, see also PM04 and Pipino, Matteucci \& Chiappini
2006, hereafter PMC06), 
the stellar
metallicity distribution, $\Upsilon_{prog}$, as the fraction of stars
formed in a given metallicity (Fe/H or Mg/Fe) bin.

The possibility to predict $\Upsilon_{prog}$ properties is important
in the context of this study, because it allows us to infer
the average abundance ratios in the stars. These ratios are useful
when we want to compare our theoretical predictions to the observations,
often given in terms of \emph{SSP-equivalent} values, namely 
luminosity-weighted measures of the properties in the stellar component (see PMC06).
The fact that we deal with old objects without any merger-induced
SF, guarantees that mass-weighted and luminosity-weighted values are
very similar in the more massive ellipticals (e.g. Arimoto \& Yoshi 1987, Matteucci, Ponzone \& Gibson 1998).

\begin{figure}
%\epsscale{.80}
\includegraphics[width=8cm,height=8cm]{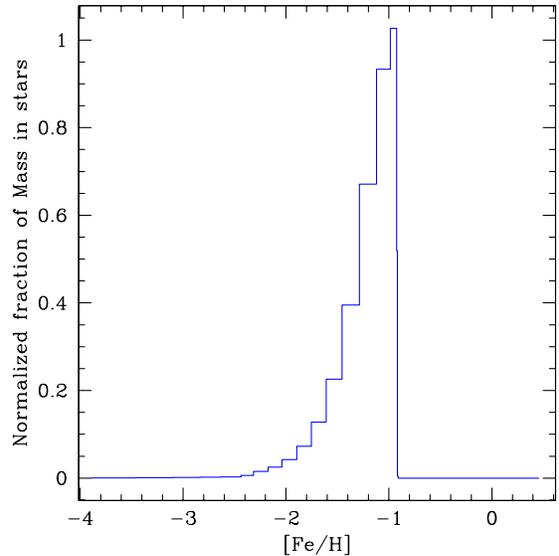}
\caption{Predicted stellar metallicity distribution $\Upsilon_{prog}$ as a function of [Fe/H] for the
progenitor A.}
\label{prog_A}
\end{figure}

\begin{figure}
%\epsscale{.80}
\includegraphics[width=8cm,height=8cm]{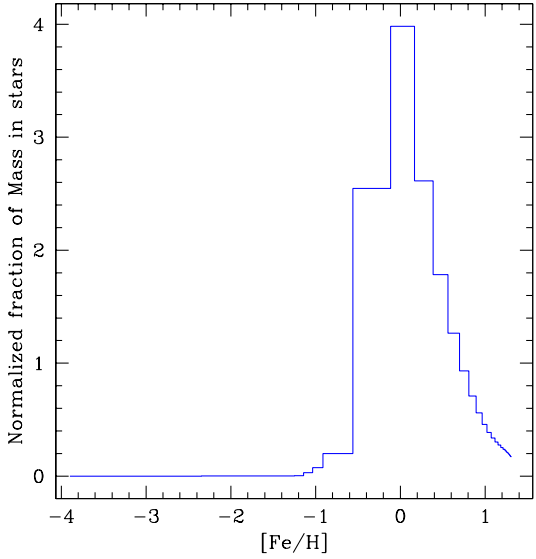}
\caption{Predicted stellar metallicity distribution $\Upsilon_{prog}$ as a function of [Fe/H] for the
progenitor B.}
\label{prog_B}
\end{figure}

The assumed galactic building blocks have the following characteristics:
\begin{itemize}
\item[A:] a $\sim 5\times 10^{8}M_{\odot}$ building block (luminous mass) which resembles a sort of present-day
dwarf spheroidal galaxy (dSph). It has been obtained by assuming a star formation efficiency 
of $\nu =0.1 \rm \, Gyr^{-1}$, while the infall timescale is $\tau =0.5 \rm \, Gyr$ 
in agreement with the prescriptions of Lanfranchi \& Matteucci (2004,2007).
For this model we predict [$<Mg/Fe>$] = 0.28 dex and [$<Fe/H>$] = -1.2 dex,
and a  stellar metallicity distribution
function $\Upsilon_{prog}$ shown in Fig.~\ref{prog_A}. The star formation in this block lasts 2.4 Gyr, which is the time at which a supernovae-driven
galactic wind occurs.

\item[A+:] a $\sim 5\times 10^{8}M_{\odot}$ building block has been obtained by assuming a  higher star formation efficiency with respect to the previous case
of $\nu =10 \rm \, Gyr^{-1}$, while the infall timescale is $\tau =0.5 \rm \, Gyr$.

For this model we predict [$<Mg/Fe>$] = 0.57 dex and [$<Fe/H>$] = -0.9 dex,
%and a $\Upsilon_{prog}$ stellar metallicity distribution
%function shown in Fig.~\ref{prog_A+}. 
The star formation here lasts only 130 Myr due to the faster occurrence of a galactic wind.

\item[B:] a $\sim 5\times 10^{8}M_{\odot}$ building block has been obtained by assuming a star formation efficiency of $\nu =1 \rm \, Gyr^{-1}$, while the infall timescale is $\tau =3 \rm \, Gyr$.

For this model we predict [$<Mg/Fe>$] = -0.1 dex and [$<Fe/H>$] = 0.63 dex,
and a stellar metallicity distribution $\Upsilon_{prog}$
function shown in Fig.~\ref{prog_B}.The star formation here lasts 1.8 Gyr.

\item[E:] a $\sim 10^{11}M_{\odot}$ elliptical galaxy which
matches both the MMR and the MFMR as well as the Colour-Magnitude Relation (CMR, Bower et al
1992). We refer to PM04's Model IIb for the same mass.
For this model we predict an overall [$<Mg/Fe>$] = 0.25 dex and [$<Fe/H>$] = 0.04 dex.
The stellar metallicity distribution is presented in Fig.~\ref{prog_AB} (solid line).
We refer the reader to PMC06 for a more detailed description of the
stellar metallicity distributions $\Upsilon_{prog}$ for such a model.

\end{itemize}

%\begin{figure}
%\epsscale{.80}
%\includegraphics[width=8cm,height=8cm]{}
%\caption{Predicted stellar metallicity distribution as a function of [Fe/H] for the
%progenitor A+.}
%\label{prog_A+}
%\end{figure}

The final massive elliptical (F) we want to simulate is a $\sim 2 \times 10^{11} M_{\odot}$ galaxy.
In particular, we expect it to have [$<Mg/Fe>$] = 0.27 dex. %and [$<Fe/H>$] = 0.08 dex,
according to PM04's Model IIb predictions for the same mass.

For the sake of simplicity we will assume that our model galaxies do not have 
radial gradient in $\alpha$-enhancement.
As shown by Pipino, D'Ercole \& Matteucci (2008), in fact, even though
most ellipticals form outside-in, the expected strong and positive [$<\alpha /Fe>$] gradient 
can be affected by the metal rich gaseous flows inside the galaxy acting togheter
with the SFR.
We recall that also osservations suggest that the observed gradient slope in the [$<Mg/Fe>$] has
a null mean value (e.g. Mehlert et al. 2003).
Therefore we will refer to a one zone model in which the metallicity and
the $\alpha$ enhancement do not vary with radius.
In general, we will expect that mergers cannot account for the steep metallicity
gradient observed in the majority of ellipticals (e.g. Carollo et al. 1993), and we
postpone to a forthcoming paper the analysis of
the gradients survival to several dry-mergers.

Under these assumptions, we predict the properties of the Composite Stellar
Populations of the merger remnant in a straightforward manner.
In fact, the stellar metallicity distribution function for the end product
of a dry-merger is simply $\Upsilon_{\rm final}$ summed over all progenitors
and can be written as:
\begin{equation}
\Upsilon_{\rm final} = \sum_{prog} M_{*,prog}\cdot \Upsilon_{prog} (t,Z)/\sum_{prog} M_{*,prog} \; .
\label{eq1}
\end{equation}
where $M_{*,prog}$ is the stellar mass of the single progenitors.
Similar equations hold for other distributions as 
functions of either [Mg/Fe] or [Fe/H].
We stress again that this is possible because we are studying dry merger
remnants, i.e.
systems where no further SF is allowed to occur.
%, otherwise we should 
%introduce another term in the right hand side of eq.~\ref{eq1}
%which takes into account the new stars formed (and their metallicity).

{ We also emphasise the fact that we present an exercise whose assumptions
are rather extreme (i.e. galaxies formed only via dry-mergers)
and without taking into account the observed merger rate
and its relations with redshift, environment and, possibly, morphology
of the progenitors (e.g. Lin et al. 2008).}

\section{Results and discussion}

\subsection{Multiple dry mergers of equal progenitors}

Let us first assume the extreme case in which our massive elliptical
has been made by merging of several progenitors of the kind \emph{A} only, as expected
from galaxy formation models which assume a short SF process
at high redshift, but let the galaxy assembly happen much later (e.g. De Lucia et al. 2006,
Kobayashi et al. 2007).

In order to have the right final mass, we need 400 of such small builiding blocks.
Since progenitor A has been built in order to yield the final correct $\alpha$
enhancement, our massive spheroid will have a Mg enhancement of 0.28 dex and will match
fairly well the average observational value for galaxies of the same mass. 
It is rather intuitive from eq.~\ref{eq1} that the final stellar metallicity distribution
will still look like Fig.~\ref{prog_A}, therefore its final
metallicity in terms of [Fe/H] will remain very low, thus not
matching either the MMR or the CMR.

The predicted value of the SFR per unit mass is 0.02/Gyr for the progenitor of kind \emph{A}.
Again, it is rather intuitive that the final spheroid will have the same value,
at variance with the results from Thomas et al. (2005), which
require this factor to be at least 2-3/Gyr, namely a factor of a hundred higher.
Such a high SFR is needed also to reproduce the observated SFR in
Lyman Break ($\sim 5-940 \, \rm h^{-2}M_{\odot}yr^{-1}$, Shapley et al. 2001) ad SCUBA (e.g. Swinbank et al. 2004) galaxies.

On the other hand, a quasi-monolithic model can have naturally  
the required SFR per unit mass. In fact the prediction PM04 (Fig.~\ref{sfr})
is in good agreement for what concerns \emph{shape, timescale and mean redshift of formation}
with those inferred by Thomas et al. (2005, see their Fig. 10),
the only difference being the sharp truncation due to the galactic wind.
It should be noticed that many models based on the hierarchical clustering, which claim to have incorporated
downsizing (e.g. De Lucia et al 2006, Kobayashi et al. 2007) have average SFR per unit mass lower
at least by a factor of 3-5 than what is required from chemical evolution studies
and line-strenght indices analysis to reproduce the [Mg/Fe] in  massive ellipticals;
according to our calculations,
with such  a low SFR per unit mass is possible to reproduce only a very modest $\alpha$-enhancement (if any).
In fact, such a value for the SFR per unit mass ($\sim 1/Gyr$) will only
suffice to explain the [$<Mg/Fe>$] $\sim$ 0.1 dex of the less massive spheroids.  

We tried to overcome the problem of having a too low SFR by introducing another ad hoc building block, namely the progenitor \emph{A+},
which has a SFR per unit mass of the order of unity.
In this case, however, the lack of agreement with the MMR is much more evident, % (Fig.~\ref{prog_A+}),
and also the predicted [$<Mg/Fe>$] is too high.
We notice in passing that if we allow for a subsequent gas-rich merger
triggering a substantial episode of SF, we may be able to reproduce the MFMR,
but we fail in obtaining other properties, such as the CMRs or the MMR, as shown by
Pipino \& Matteucci (2006) (see, e.g., the models discussed in their Sec. 3.3).

\begin{figure}
%\epsscale{.80}
\includegraphics[width=8cm,height=8cm]{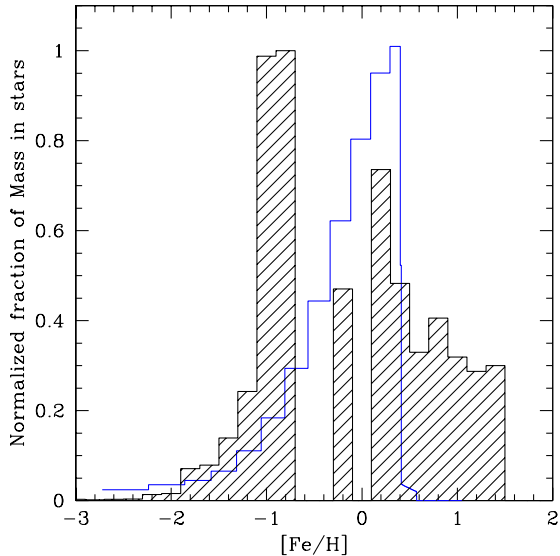}
\caption{Shaded hystogram: p redicted stellar metallicity distribution $\Upsilon_{final}$ as a function of [Fe/H] for the
case in which the final massive galaxy is made 50\% of progenitor A-like building blocks
and the rest of progenitors B. For comparison, the prediction for the $\Upsilon_{prog}$ of the PM04 best model (case E) is
presented (solid line).}
\label{prog_AB}
\end{figure}

\begin{figure}
%\epsscale{.80}
\includegraphics[width=8cm,height=8cm]{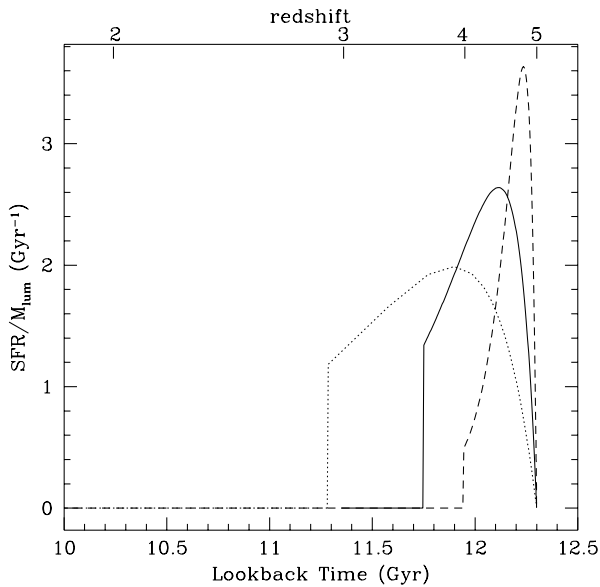}
\caption{Predicted SFR per unit mass for several masses, PM04's best model.
The model E is represented by a solid line.}
\label{sfr}
\end{figure}

On the other hand, we can have the extreme case in which
the galaxy is created by several progenitors of the kind \emph{B}. The results of this analysis show that this model can reproduce
the MMR but it predicts  an underabundance of $\alpha$-elements relative to Fe, 
at variance with observations.

We conclude that we cannot form massive spheroids from a sequence of
several dry-mergers between building blocks of the same kind (similar mass and chemical properties),
even if the progenitors are chosen to have the correct $\alpha$-enhancement.
For the same reason, present-day low mass ellipticals which satisfy both the MMR
and the MFMR cannot be the building blocks of massive ones. 
This conclusion can be extended to the CMR, since the colour differences
are mainly driven by metallicity.
Remarkably, similar conclusions have been obtained by Ciotti et al. (2007)
by studying the dynamical properties of ellipticals.
Therefore, pure dissipationless
merger of similar stellar systems \emph{cannot change the metallicity, the $\alpha$-enhancement, the colours
and the virial velocity dispersion.}

\subsection{Multiple dry mergers of different progenitors}

Now we relax the extreme assumption of the previous section and allow for two
or more kind of progenitors for our massive elliptical.
If, for simplicity sake, we have a fraction $f=$50\% of the final mass coming from progenitors
of the type A and 1-$f=$50\% from progenitors B,
the final stellar metallicity distribution (shaded hystogram in Fig.~\ref{prog_AB}) will be closer to the one
expected for a normal elliptical (solid line in Fig.~\ref{prog_AB})and the outcome will match the CMR and the MMR, being its
final [$<Fe/H>$] = 0.66 dex, but the predicted [$<Mg/Fe>$] = 0.06 dex is still too low.
Moreover, this scenario cannot represent a solution for the still too low SFR per unit mass.

If we repeat the same exercise with model \emph{A+} and B,
these latter quantities get in a better agreement with the values
inferred from the observations, and
we also notice an improvement for the [$<Mg/Fe>$] which now amounts to 0.1 dex.

We admit that the portion of the parameter space that we are investigating is quite
small - although the choice of the  models is sensible - the main aim of this investigation
being the study of a few clear and extreme cases.
Such examples serve to probe to which extent the random nature of
the merger process models can be accomodated within the observational
uncertainties.
A more comprehensive analysis featuring a proper merger history based on the
hierachical growth of structure and a self-consistent chemical evolution 
is in preparation (Pipino et al, 2008).

However, even in the case in which either two more suitable progenitors can be found, or
a different mixture of several progenitors can predict the right final chemical
properties for a given final galactic mass, several questions arise: i) why are only dSph (i.e. progenitor A) still
observable in the local universe? (but see Robertson et al. 2005).
ii) why is the fraction $f$ such that none of the two classes is predominant?
iii) since the [$<Mg/Fe>$] correlates with the final galactic mass, one should expect
progenitors with different initial (i.e. pre-merger) properties - which scale accordingly to the final mass of the object
- to live in the early universe. How it is possible that
they know in advance what they are about to build later on? 
%{ mi sembra di ricordare che i gerarchici hanno una spiegazione per questo, prova a chiedere..}
Finally, even if a selection mechanism is at work and it leads to an agreement
between model and observed chemical properties, it must be able
to account for other scaling relations, such as the Faber-Jackson (Faber \& Jackson, 1976) and the Kormendy (1977) relations
as well as the Fundamental Plane (e.g. Dressler et al. 1987).

{ It is also interesting to notice that Bournaud et al. (2007) claimed that
repeated minor mergers - as the ones studied in this section - can theoretically form massive boxy elliptical galaxies
without major mergers, being more frequent than the latter, in particular at moderate redshifts. 
The mechanism put forward by Bournaud et al. (2007) could explain the morphology and the dynamical properties
of the merger remnant; in particular it might be a viable alternative to overcome the issues 
in the major-merger scenario (Naab \& Ostriker, 2007)
in order to explain the high boxiness of massive ellipticals. Unfortunately,
Bournaud et al. (2007) explored only too a narrow mass range to understand
whether their argument helps in reproducing the observed scaling relations
for elliptical galaxies.}

\subsection{Multiple minor-dry merger on an already formed elliptical galaxy}

At variance with the previous section, we now test the scenario in which the final 
galaxy is built via a series of a minor dry merger, namely adding 
several progenitors of either type (A-like or B-like) to a galaxy like progenitor
E until we double its mass (therefore we need roughly 200 of small building blocks).
Since the model E stars contain the right amount of $\alpha$-enhancement and mean metallicity,
this case will help us in assessing whether the accretion of several progenitors of either type A or B
can worsen the agreement with observations.
Following the same line of reasoning of the previous sections, we first assume
that we want to build the final galaxy as the sum of E and only type A+ progenitors.
We obtain the $\Upsilon_{\rm final}$ for model F according to eq.~\ref{eq1} and 
we find that the final [$<Mg/Fe>$] = 0.26 is not exceeding the observational boundaries.
If we assume that F is the result of a progenitor like E plus roughly
two hundreds of building blocks of the type B, we predict [$<Mg/Fe>$] = 0.13,
which is on the lower observational boundary (see fig.~\ref{fig_espl}, upper panel).
The two cases presented above bracket a region of the paramenter
space in which the final galaxy F can be obtained through a series
of dry mergers involving a galaxy like E and progenitors like A, A+ and B
in different mixtures.
Three main conclusions can be drawn on a scenario featuring multiple small
progenitors accreted by a medium-sized ellipticals until it double its
mass:
i) multiple minor dry mergers cannot be ruled out on the basis 
of current observations, but they are only small perturbations
for a massive elliptical which formed monolithically;
ii) they may explain the observed scatter in the [$<Mg/Fe>$] values
at a given galactic mass, although it can also be explained in the framework
of the revised monolithic scheme by small differences in either the
star formation efficiency or the infall timescale with respect
to the PM04 best value (tuned to represent the average galaxy);
iii) in any case, they cannot explain the trend of [$<Mg/Fe>$] with
$\sigma$, because, even if they may lead to a modest increase
in [$<Mg/Fe>$], the stellar
velocity dispersion does not increase (Ciotti et al., 2007).

It is obvious that a few events like the ones depicted in this section
can occur, for instance, in a dense environment such
as a cluster of galaxy 
where also a residual on-going SF is detected
in massive ellipticals and cD (Bildfell et al. 2008).
The amount of SF inferred from UV spectra (Kaviraj et al. 2007)
from redshift 1 to the present-day ({ and consistent with the merger rate by Khochfar \& Burkert, 2006})
will lead to a modest increase (1-5 percent) in the stellar mass, even tough
no SNII explosions have been detected to-date (Mannucci et al. 2007, but they might
still occur in S0, e.g., Pastorello et al. 2007);
therefore  we consider it either as an \emph{accident} or a consequence of the
environment, rather than a signature of any particular galaxy formation scenario.
In fact, according to our previous calculations (Pipino \& Matteucci 2006),
such a low intensity late SF episode cannot significantly lower
the [$<Mg/Fe>$] ratio.

Interestingly, recent observational evidences (Daddi et al. 2005, Trujillo et al. 2007) favour an increase in the size ellipticals
by a factor of about 4 since redshift z=1, plausibly associated with the occurrence of dry-mergers.
However, the simulations show that this kind of accretion occurs mainly outside one effective radius
(Naab et al, 2007) and it might be due to accretion of small satellites (Daddi et al. 2005);
therefore we do not expect them to affect the properties
of the galactic core, whose stars obey to the MMR and the MFMR.
Again, even in the framework of the revised-monolithic scenario, 
these episodes are unavoidable either in a dense environment or 
considering the fact the massive ellitpicals are 12 Gyr old, therefore they
had enough time to interact with their satellites.

\subsection{Late major dry merger}

Finally, we want to test the feasibility of a major dry merger
between two massive spheroids of the kind \emph{E} in order to produce the galaxy \emph{F}.
In this case, the final elliptical will double its mass and keep
a mean [$<Mg/Fe>$] = 0.25 dex, which does not differ much from the value
expected by model F.
Nevertheless, a problem arises from the fact that, for a pure dissipationless merger
between two objects of equal mass and same velocity dispersion $\sigma$,
the final object will double its mass, but preserve $\sigma$ (Ciotti et al. 2007). 
Therefore, we cannot move along the direction of the observed
[$<Mg/Fe>$]-$\sigma$ relation. This seems to be the case also when 
the stellar central velocity dispersion is allowed to (modestly) increase 
due to non-homology effects (Nipoti et al., 2003).

We stress that this clear and straightforward consequence
of the virial theorem is often neglected in works
which aim at reconciling the prediction from the hierarchical
clustering scenario with the evidences coming from the chemistry.

Nonetheless, we can exploit the [$<Mg/Fe>$]-mass relation to infer some
constraints on the number of major dry-mergers involving massive
spheroids. From Thomas et al. (2005), we know that 
$[<Mg/Fe>]=-0.459+0.062 log(M_*)$ with an intrinsic scatter of $\pm$0.05 dex.
This means that an elliptical galaxy which satisfies the average trend,
can undergo either $\sim$ 2-4 major dry-mergers (if only one has mass ratio 1:1, the rest being either 1:2 or 1:3),  
or $\sim$ 2-3 major dry-mergers (if the mass ratio is always 1:1, i.e. at the first step we create two galaxies, each one via a merger
of two units like model E (2xE), and then we let them merge togheter), before crossing the boundary set by observations.
{ Remarkably, such a limit is in agreement with recent observational estimates for the avearge
number of major mergers experienced by elliptical galaxies since redshift 1.2 (e.g. Lin et al. 2008).
Unfortunately, the nature of our exercise does not allow us to either put precise constraints
or to make predictions on the merger rate. However, if we take into account the fact that
the dry-merger rate seems to decline quite strongly with redshift, we may 
argue that the maximum major-merger rate allowed by the MFMR is 0.2-0.3/Gyr 
in agreement with Bell et al. (2006)'s estimate.}
In any case, following the same argument of the previous sections,
no arbitrary sequence of dry-mergers of a model E galaxy 
with similar progenitor can ever form a $10^{12}M_{\odot}$
spheroid with the highest observed [$<Mg/Fe>$] for that mass. 

\section{Conclusions}

\begin{figure}
%\epsscale{.80}
\includegraphics[width=8cm,height=10cm]{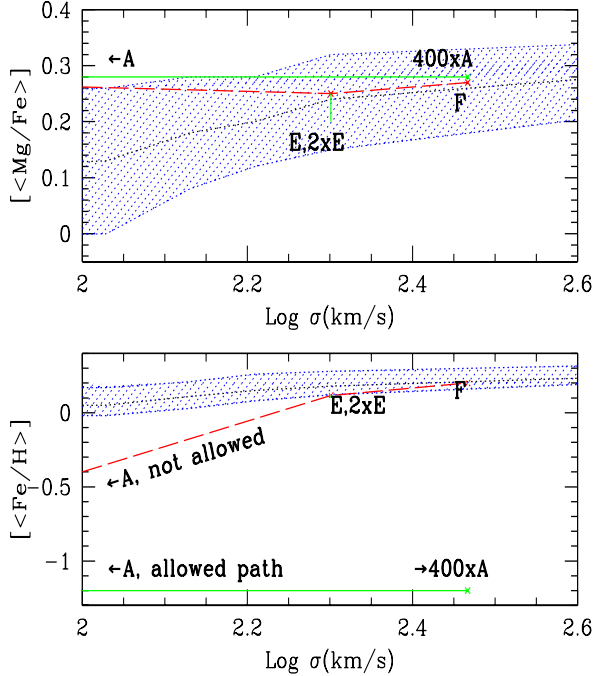}
\caption{Predicted evolutionary paths (solid: allowed by the analysic of the chemical properties;  dashed: not allowed, see text) 
in the [$<Mg/Fe>$]-$\sigma$ (upper panel) and [$<Fe/H>$]-$\sigma$ (lower panel) planes.
The initial, intermediate and final points are marked according to the galaxy (progenitor) type
and the merger end-product, respectively.
The shaded areas encompass the observations (Nelan et al. 2006),
whose average trends with the velocity dispersion are given by the black dotted lines.}
\label{fig_espl}
\end{figure}

Before drawing our final remarks, we summarize the results
obtained in the studied case by means of fig.~\ref{fig_espl}, where the 
average [$<Mg/Fe>$] and the average [$<Fe/H>$] are plotted versus $log(\sigma)$.
In this figure we sketch the allowed paths as solid lines, 
whereas we plot as dashed lines what is impossible on the basis
of stellar population arguments.
We compare them with the observations by Nelan et al. (2006):
the black dotted line being the mean value,
and the shaded area brackets the observational scatter reported
by the authors.
These values are in agreement with the previous analysis
by Thomas et al. (2005), although in Nelan et al. (2006) 
the slope of the MFMR is somewhat flatter at high velocity dispersions.

As done in the previous sections, we follow the merger history which
links progenitors \emph{A} to the \emph{intermediate stage E}
to the \emph{massive elliptical F} as an example of a possible accretion history
which may lead to the creation of a massive spheroid from small subunits.
With \emph{400xA} we name the outcome of the multiple mergers
described in Sec. 3.1. We already know that some of the chemical properties of
such a galaxy will differ from those featured by model F.
We assume that the progenitors of type A inhabit the upper (lower) left
corner in the upper (lower) panel, without any further specification
on their initial $\sigma$ and on the dynamical outcome of the multiple mergers.
To be conservative, in fact, we assume that they can somehow
increase their $\sigma$ \footnote{This might happen if the progenitors
are already bound to each other, as in
they case in which they inhabit the same massive Dark Matter halo.
However, the energy must have the exact value for the merger end-product 
to obey to the Fundamental Plane, therefore this case seems rather unlikely.}, 
%moreover such a system should collapse on a very short timescale
but we recall that, were the build-up of model
400xA a sequence of \emph{pure parabolic mergers\footnote{This
seems to be the case for nearly 50\% of the merger events (Khochfar \& Burkert, 2006)} between
collisionless systems}, its final $\sigma$ would
be equal to the maximum value of the stellar velocity dispersion
among the progenitors (Ciotti et al., 2007, see also Nipoti et al. 2003
for the effects of non-homology). This means
that the remnant merger would probably lie very close to A in the upper panel
of fig.~\ref{fig_espl}, without moving leftward as we assume in our
simple sketch.
From the upper panel of  Fig.~\ref{fig_espl}, we notice
that the model 400xA is fairly close to the model that we want to reproduce
through the sequence of mergers (F), therefore
we mark this path (solid line) in the [Mg/Fe]-$\sigma$ plane as \emph{allowed},
meaning that, on pure chemical evolution basis, it is possible
to create a massive sphoroid which satisfies the MFMR starting from
$\alpha$-enhanced building blocks.
On the other hand, the lower panel of Fig.~\ref{fig_espl}, tells
us that the galaxy of type 400xA falls short in reproducing the MMR by more than
two orders of magnitude. A path joining A with E and F (dashed line)
is impossible if we want to satisfy both the observed relationships.
We do not show here the paths leading to a final F galaxy from building blocks
of type B. It is intuitive that they will be \emph{allowed} 
on the metallicity-$\sigma$
plane, but they will be outside the region delimited by the observations in the
[Mg/Fe]-$\sigma$ plane.
If we allow galaxies to form through a sequence of mergers involving
different progenitors, given the random nature of the hierarchical assembly it is natural 
to expect a scatter in the predicted MMR and the MFMR much larger
than the observed ones and no slope at all.
We expect this to happen also in the most recent models based
on the hierarchical assembly and even if the features
several gestures toward the \emph{downsizing},
as it has indeed been shown by Nagashima et al. (2005).

This fact and our previous works on the quasi-monolitic formation
of massive ellipticals (PM04) as well as  the effect of major wet mergers 
(Pipino \& Matteucci, 2006) allow us to derive the first important conclusions:

\emph{i) The diagnostic power of the MMR and MFMR relies in the fact
that the mechanisms required to satisfy the former tend to worsen
the agreement with the latter, and viceversa}.

\emph{ii) Only when most of the star formation process and the galactic assembly occur
at roughly the same time and the same place both relations can be fulfilled.}

As expected from the discussion of section 3.4, the outcome (called \emph{2xE} in Fig.~\ref{fig_espl})
of a major dry merger involving two massive ellipticals,
will occupy the same place of its progenitor \emph{E} and does not
move to where it is expected (i.e. at \emph{F}), therefore we mark
this path as \emph{not allowed}
%Even if we neglect dynamics and assume that a dry-merger is a viable channel to reproduce
%the most massive ellipticals with the lowest $\alpha$-enhancement (because
%the path will be parallel to the abscissa axis),
%it is still to be understood which are the progenitors of the spheroids with the 
%highest [$<Mg/Fe>$]. 

Moreover, the model \emph{E} has been designed to represent an average
elliptical of medium size, although the observations tell us that there are
galaxies of the same mass but lower $\alpha$-enhacement (as low as 
0.1 dex, lower boundary of the shaded region  - Fig.~\ref{fig_espl}, upper panel).  
This means that, in principle, we should be able to observe also giant
ellipticals with [$<Mg/Fe>$] as low as $\sim$ 0.1 dex.
Since this is not the case, we have not only the need of a way to increase
$\sigma$ without having any substantial wet mergers (and further SF),
but also we need to invoke some selective mergers.
Other constraints will come from the study of the joint evolution of
the MFMR and MMR with redshift.
%To date, in fact, the MFMR does not seem to evolve in time, at least up to redshift
%0.4 (Ziegler et al., 200?), therefore the dry mergers should happen
%at earlier epochs.

In conclusion, it seems hard to reproduce all giant ellipticals
either via a \emph{pure} sequence of multiple minor dry mergers, or via  major dry mergers.
However, the scatter of the MFMR is such that 
the occurrence of 1-3 major-dry mergers during the galactic lifetime 
cannot be ruled out { and indeed it is in agreement with the observations
(e.g. Bell et al., 2006)}. 
For the same reason, several accretion episodes of small
satellites onto a massive 'monolithic' elliptical galaxy,
can be classified as perturbations which may help explaining
the observational scatter.
{ Moreover, it is still to be understood whether a mixed (namely including
both wet- and dry-mergers) model which tracks the observed evolution
of the merger rates amongst different galactic morphologies (e.g. Lin et al., 2008) 
can overcome the above difficulties. \footnote{Thomas \& Kauffamann (1999) already showed that this is not the case for
the class of earlier hierarchical models not taking into account the \emph{downsizing}.}
This is beyond the scope of the present paper, which (along with PM06)
aims at being only a first step into a quantitative analysis
of the chemical properties of merger remnants.
We stress that we present some rather extreme cases (i.e. galaxies formed only via dry-mergers)
as opposed to PM06 where we presented only wet-mergers.
Results by a semi-analytical model which incorporates
several gestures toward the reproduction of the \emph{downsizing}
as well as the galactic colour bimodality,
with a full and self-consistent treatment of the chemical evolution 
will be presented in a forthcoming paper (Pipino et
al., in preparation).}

%\acknowledgments
\section*{Acknowledgments} 
We acknowledge useful discussions with T. Naab.
We thank the referee for the careful reading.
L. Ciotti is warmly thanked for a timely reading of the paper
and many enlightening comments.

%Maiolino's grant?

\clearpage

\end{document}